\documentclass{article}
\usepackage{spconf,amsmath,graphicx}

\usepackage{enumitem}
\setlist{nosep, leftmargin=14pt}

\usepackage{mwe} 
\usepackage{multirow}
\usepackage{amssymb}
\usepackage{algorithmic}
\usepackage{algorithm}
\usepackage[colorlinks=true, urlcolor=blue, linkcolor=red]{hyperref}
\usepackage{color, colortbl}

\definecolor{Worse}{rgb}{0.9,0.9,1}
\definecolor{Better}{rgb}{1,0.9,0.9}

\title{Adaptive Aggregation Weights for Federated Segmentation of Pancreas MRI}
\name{\begin{tabular}{c}Hongyi Pan$^1$, Gorkem Durak$^1$, Zheyuan Zhang$^1$, Yavuz Taktak$^2$, Elif Keles$^1$, Halil Ertugrul Aktas$^1$, \\Alpay Medetalibeyoglu$^3$,
Yury Velichko$^1$, Concetto Spampinato$^4$, Ivo Schoots$^5$, Marco J. Bruno$^6$, \\Rajesh N. Keswani$^7$, Pallavi Tiwari$^8$, Candice Bolan$^9$, Tamas Gonda$^{10}$, Michael G. Goggins$^{11}$, \\Michael B. Wallace$^9$, Ziyue Xu$^{12}$, Ulas Bagci$^1$\end{tabular}
}
\address{$^1$Department of Radiology, Northwestern University, Chicago, IL, USA\\
$^2$Department of Radiology, Istanbul Faculty of Medicine, Istanbul University, Istanbul, Turkey\\
$^3$Department of Internal Medicine, Istanbul Faculty of Medicine, Istanbul University, Istanbul, Turkey\\
$^4$Department of Electrical, Electronic and Computer Engineering, University of Catania, Catania, Italy \\
$^5$Department of Radiology and Nuclear Medicine, Erasmus Medical Center, Rotterdam, Netherlands\\
$^6$Department of Gastroenterology and Hepatology, Erasmus Medical Center, Rotterdam, Netherlands\\
$^7$Department of Gastroenterology and Hepatology, Northwestern University, Chicago, IL, USA\\
$^8$Department of Biomedical Engineering and Radiology, University of Wisconsin, Madison, WI, USA\\
$^9$Division of Gastroenterology and Hepatology, Mayo Clinic, Jacksonville, FL, USA\\
$^{10}$Division of Gastroenterology and Hepatology, New York University, New York, NY, USA\\
$^{11}$Division of Gastroenterology and Hepatology, Johns Hopkins University, Baltimore, MD, USA\\
$^{12}$NVIDIA, Bethesda, MD, USA
}
\begin{document}
\ninept
\maketitle
\begin{abstract}
Federated learning (FL) enables collaborative model training across institutions without sharing sensitive data, making it an attractive solution for medical imaging tasks. However, traditional FL methods, such as Federated Averaging (FedAvg), face difficulties in generalizing across domains due to variations in imaging protocols and patient demographics across institutions. This challenge is particularly evident in pancreas MRI segmentation, where anatomical variability and imaging artifacts significantly impact performance. In this paper, we conduct a comprehensive evaluation of FL algorithms for pancreas MRI segmentation and introduce a novel approach that incorporates adaptive aggregation weights. By dynamically adjusting the contribution of each client during model aggregation, our method accounts for domain-specific differences and improves generalization across heterogeneous datasets. Experimental results demonstrate that our approach enhances segmentation accuracy and reduces the impact of domain shift compared to conventional FL methods while maintaining privacy-preserving capabilities. Significant performance improvements are observed across multiple hospitals (centers).
\end{abstract}
\begin{keywords}
Federated learning, adaptive aggregation weights, pancreas segmentation, magnetic resonance imaging
\end{keywords}
\section{Introduction}
Magnetic Resonance Imaging (MRI) has become an essential tool in medical imaging, particularly for pancreas volumetry (requiring segmentation), pancreatic cancer risk estimation, and treatment planning~\cite{cai2016pancreas,kumar2019automated,cai2019pancreas,yao2020advances,zhang2023tu1986}. 
The accurate segmentation of the pancreas from MRI scans is fundamental for quantitative analyses that inform clinical decision-making and personalized treatment strategies. However, developing robust deep-learning models for pancreas segmentation presents significant challenges. The pancreas exhibits a highly complex and variable anatomical structure, which complicates creating models that can consistently delineate its boundaries across diverse patient populations. Additionally, variability in MRI scanner hardware, imaging protocols, and patient demographics introduces a phenomenon known as domain shift, where models trained on data from one institution or imaging setup perform poorly when applied to data from different sources~\cite{cai2016pancreas,kumar2019automated}.


To address these challenges, alternative strategies such as federated learning (FL) have gained traction, allowing institutions to collaboratively train models without sharing raw data~\cite{li2020review,zhang2021survey,guo2022auto,jiang2023fair,gholami2024digest,miao2024contrastive,pan2024frequency}. This decentralized approach is particularly advantageous in medical imaging, where privacy regulations, restrict the transfer of sensitive patient data~\cite{topaloglu2021pursuit}. However, despite its promise, FL faces significant challenges in medical applications due to the inherent heterogeneity of data across institutions. Variability in imaging protocols, scanner resolutions, and patient populations introduces domain shifts that can impair model generalization.

These challenges are especially evident in pancreas MRI segmentation, which has never been performed under the FL setting. Besides, the pancreas is a small and anatomically variable organ, making it difficult to segment across datasets consistently. Available algorithms are focused on CT scans, and MRI based algorithms are absent. Regarding conventional FL methods,  Federated Averaging (FedAvg)~\cite{mcmahan2017communication,collins2022fedavg} and its variant FedProx~\cite{li2020federated}, aggregate models uniformly by assuming homogeneity among clients. This assumption can degrade model performance, as some clients may contribute less reliable updates due to differences in local data distributions. To address this issue, adaptive aggregation strategies~\cite{tan2022adafed,zhang2023federated,zhang2023fedala,chen2024fedawa} have been introduced to weigh client contributions based on their local models' performance and relevance to the global model.

\begin{figure*}[tb]
\centering
\includegraphics[width=.7\linewidth]{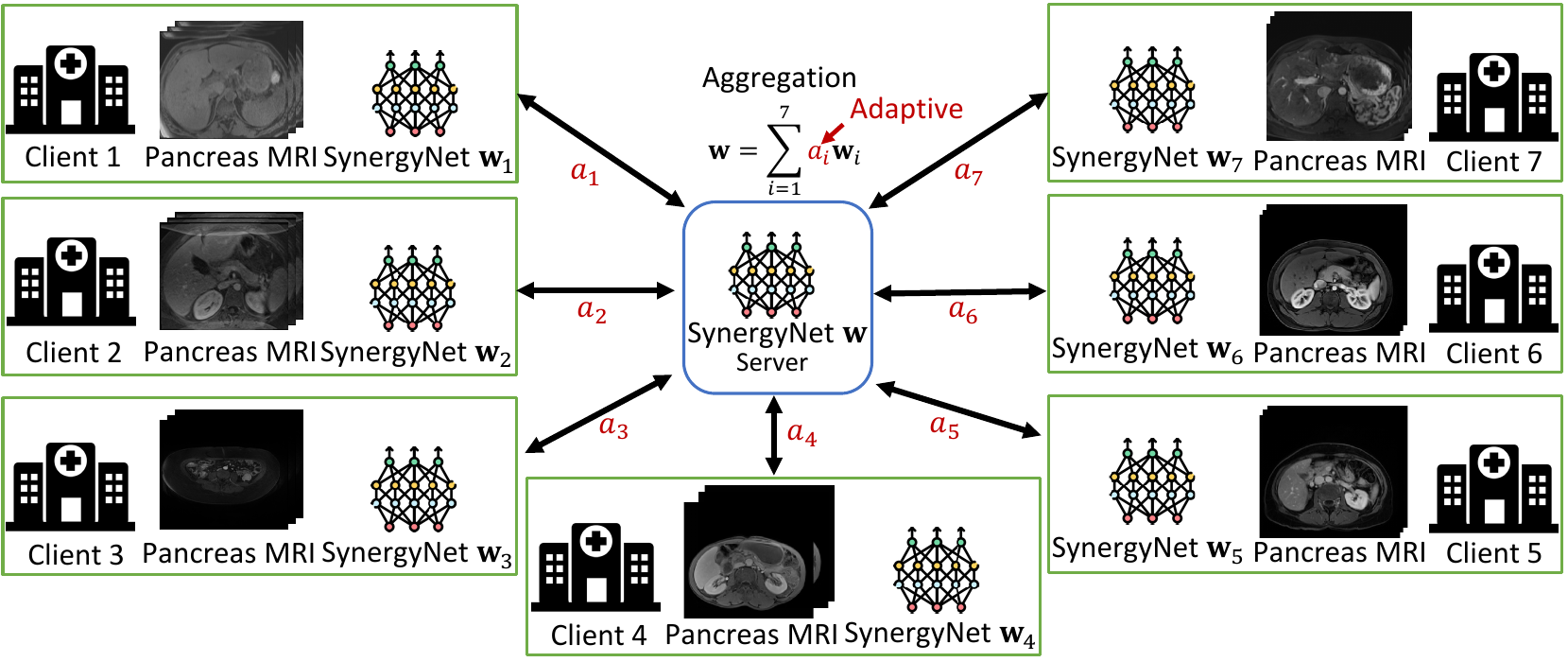}
\caption{Adaptive aggregation weights for federated pancreas MRI segmentation. Unlike the traditional FedAvg algorithm, where aggregation weights remain fixed, our method updates weights dynamically based on the validation loss gap before and after aggregation, ensuring that clients contributing positively to model performance are given more weight in subsequent rounds.}\vspace{-10pt}
\label{fig: adaptive aggregation weights}
\end{figure*}

In this paper, we comprehensively evaluate FL algorithms for pancreas MRI segmentation. Building upon this evaluation, we propose a novel approach that leverages adaptive aggregation weights to improve FL performance. By dynamically adjusting the contribution of each client during model aggregation, our method enhances the model's ability to generalize across diverse institutions as Fig.~\ref{fig: adaptive aggregation weights}. Unlike traditional FL methods, which assume equal contributions from all clients, our approach customizes the aggregation process based on local domain shifts. This leads to more robust and accurate segmentation performance, addressing the inherent heterogeneity in real-world medical imaging datasets.

\section{Methodology}
In this section, we will first formulate the FL problem. Then, we will review the Federated Averaging (FedAvg) algorithm, which is widely used in federated settings. Finally, we will introduce our approach for updating the aggregation weights dynamically to enhance the FedAvg algorithm, as illustrated in Fig.~\ref{fig: adaptive aggregation weights}.

\subsection{Federated Learning Problem Formulation}
FL facilitates collaborative model training across multiple institutions without sharing raw data. Consider a central server with $K$ clients, where each client has a training dataset denoted as $\Psi_i=\{\psi_{i,0}, \ldots, \psi_{i,n_i-1}\}$ containing $n_i$ samples. The loss of model weights $\mathbf{w}$ on the $i$-th client is defined as the average loss over all samples in $\Psi_i$:
\begin{equation}
    \mathcal{L}(\mathbf{w}, \Psi_i) = \frac{1}{n_i}\sum_{j=0}^{n_i-1}\mathcal{L}(\mathbf{w}, \psi_{i,j}),
\end{equation}
where $\mathcal{L}(\mathbf{w}, \psi_{i,j})$ computes the loss on $ \psi_{i,j}$ using model weights of $\mathbf{w}$.
The server's objective is to leverage the data from all clients to identify an optimal model that minimizes the average loss across clients. Consequently, we aim to find a single model that minimizes the weighted average of client losses based on the number of samples:
\begin{equation}
    \underset{\mathbf{w}}{\text{minimize}}\sum_{i=0}^{K-1}n_i\mathcal{L}(\mathbf{w}, \Psi_i),
\end{equation}

Due to privacy constraints, clients are prohibited from sharing their local data with other clients in FL. Instead, each client uploads only its locally optimized model weights to the central server, where these weights are aggregated to update the global model. This process ensures that sensitive data remains decentralized, addressing privacy concerns while still enabling collaborative model training.

\subsection{Background: Federated Averaging}
FedAvg~\cite{mcmahan2017communication} is the most widely used FL algorithm, which works by distributing a global model to local clients, who then update it using their private data. After several local updates, the clients send their model updates (gradients or weights) back to a central server. The server then aggregates the updates by computing a weighted average, which is typically based on the number of data samples each client has, forming new global model weights $\mathbf{w}^{t+1}$:
\begin{equation}
    \mathbf{w}^{t+1}=\sum_{i=0}^{K-1}\frac{n_i}{\sum_{j=0}^{K-1}n_j}\mathbf{w}_i^{t},
\end{equation}
where, $\mathbf{w}_i^{t}$ are the model weights updated from client $i$ at round $t$.

However, FedAvg assumes that client data is independently and identically distributed (IID), which is rare in medical applications like pancreas MRI segmentation. Domain shifts between institutions due to different imaging protocols, patient populations, or scanner types can result in suboptimal generalization of the global model. To address these challenges, we propose a modified aggregation strategy that dynamically adapts to client-specific data distributions in the following section.

\subsection{Adaptive Aggregation Weights for FedAvg}
Inspired by the FedDG-GA algorithm~\cite{zhang2023federated}, to account for data heterogeneity across clients, we introduce adaptive aggregation weights into the FedAvg framework. Our approach assigns a unique weight to each client based on their local data characteristics, including performance and domain shifts. This adaptive weighting mechanism ensures that clients with more relevant or higher-quality data contribute more to the global model, while clients with outlier data have less influence:
\begin{equation}
    \mathbf{w}^{t+1}=\sum_{i=0}^{K-1}a_i^t\mathbf{w}_i^{t},
\end{equation}
where, $a_i^t$ is the adaptive aggregation weight at round $t$. The aggregation weights are initialized the same as the one in the vanilla FedAvg algorithm:
\begin{equation}
a_i^0=\frac{n_i}{\sum_{j=0}^{K-1}n_j},
\end{equation}
and they are updated based on the validation loss gap observed before and after aggregation:
Let $P_i = \mathcal{L}(\mathbf{w}_i^t, \Phi_i)$ and $Q_i = \mathcal{L}(\mathbf{w}^t, \Phi_i)$ stand for the validation loss on validation dataset $\Phi_i$ before and after model aggregation, respectively, where $\mathcal{L}$ denotes the loss function. We first compute the loss gap:
\begin{equation}
    G_i=Q_i-P_i, i=0, \dots K-1.
\end{equation}
Then, we update the aggregation weights based on the normalized loss gap:
\begin{equation}
    a_i^{t+1} = a_i^t+\frac{G_i\cdot s}{\max(|G_0|, \dots, |G_{K-1}|)}, i=0, \dots K-1.
\end{equation}
Here, $s$ is the step size for aggregation weights updating. In this work, we dynamically set $s$ at epoch $t$ as $0.1(1-\frac{t}{T})$, training the model for a total of $T$ epochs. This strategy allows the aggregation weights to change significantly at the beginning of the training process, gradually stabilizing as training progresses
To ensure the aggregation weights remain within the range $[0, 1]$
and that their sum equals to $1$, we clip the aggregation weights to the interval $[0, 1]$ and normalize them by dividing by their sum. In this approach, if a client contributes to better convergence of model weights--specifically if its corresponding validation loss before aggregation is smaller than that after aggregation--we increase its associated aggregation weights. Conversely, if a client detracts from model weight convergence--indicated by its validation loss before aggregation being worse than that after aggregation--we decrease its corresponding aggregation weights. Algorithm~\ref{al: fed} outlines the process of incorporating adaptive aggregation weights into the FedAvg algorithm.

\begin{algorithm}[tb]
	\caption{Federated averaging with adaptive aggregation weights across $K$ centers.}
	\label{al: fed}
	\begin{algorithmic}[1]
		\renewcommand{\algorithmicrequire}{\textbf{Input:}}
		\renewcommand{\algorithmicensure}{\textbf{Output:}}
		\REQUIRE Global model initial wights $\mathbf{w}^0$, training datasets $\Psi_i$ for $i=0, 1, \dots, K-1$, where the $i$-th dataset contains $n_i$ images, validation datasets $\Phi_i$ for $i=0, 1, \dots, K-1$. \\\textbf{Hyperparameters:} number of epochs $T$, adaptive aggregation weights step size $s$.
		\ENSURE  Well-trained global model wights $\mathbf{w}^T$.
            \STATE $a_i^0=\frac{n_i}{\sum_{j=0}^{K-1}n_j}$.
		\FOR{$t=0, 1, \dots, T-1$} 
            \FOR{$i=0, 1, \dots, K-1$} 
            \STATE Assign weights to local model: $\mathbf{w}_i^t=\mathbf{w}^t$;
            \STATE Update $\mathbf{w}_i^t$ using $\Psi_i$;
            \ENDFOR 
             \STATE Aggregate models: $\mathbf{w}^{t+1}=\sum_{i=0}^{K-1}a_i^t\mathbf{w}_i^{t}$;
            \FOR{$i=0, 1, \dots, K-1$} 
            \STATE Compute the validation loss before model aggregation: $P_i = \mathcal{L}(\mathbf{w}_i^t, \Phi_i)$;
            \STATE Compute the validation loss after model aggregation: $Q_i = \mathcal{L}(\mathbf{w}^t, \Phi_i)$;
            \STATE Compute the validation loss gap: $G_i=Q_i-P_i$;
            \ENDFOR 
            \FOR{$i=0, 1, \dots, K-1$} 
            \STATE Update the weight: $\tilde{a}_i^{t+1} = a_i^t+\frac{G_i\cdot s}{\max(|G_0|, \dots, |G_{K-1}|)}$;
            \STATE Clip the weight: $\tilde{a}_i^{t+1} = \min(\max(\tilde{a}_i^{t+1}, 1), 0)$;
            \ENDFOR 
            \FOR{$i=0, 1, \dots, K-1$} 
            \STATE Normalize the weight: $a_i^{t+1} = \frac{\tilde{a}_i^{t+1}}{\sum_{j=0}^{k-1} \tilde{a}_j^{t+1}}$;
            \ENDFOR 
            \ENDFOR \\
	\end{algorithmic}
\end{algorithm}

\section{Experimental Results}
This work is implemented using the extended version of the pancreas MRI dataset introduced in \cite{zhang2025large}. The dataset includes 723 T1-weighted and 738 T2-weighted MRI images, along with corresponding segmentation masks, collected from 7 centers: New York University Langone Health (NYU), Mayo Clinic Florida (MCF), Northwestern University (NU), Allegheny Health Network (AHN), Mayo Clinic Arizona (MCA), Istanbul University Faculty of Medicine (IU), and Erasmus Medical Center (EMC). The number of images available from each center is listed in Table~\ref{tab: Distribution}. Samples of the MRI slices are shown in Fig.~\ref{fig: adaptive aggregation weights}. We split the dataset into 80\% for training and 20\% for testing. Data is available at OSF\footnote{\url{https://osf.io/74vfs/}}. The experiments are implemented using PyTorch~\cite{paszke2019pytorch}.

\begin{table}[tb]
\caption{Pancreases Data Distribution.}
\begin{center}
\begin{tabular}{l|cc}
\hline
\multirow{2}{*}{\textbf{Data Centers}}&\multicolumn{2}{c}{\textbf{Modalities}}\\
&T1&T2\\
\hline
New York University Langone Health (NYU)&162&162\\
Mayo Clinic Florida (MCF)&148&143\\
Northwestern University (NU)&206&207\\
Allegheny Health Network (AHN)&17&27\\
Mayo Clinic Arizona (MCA)&25&23\\
Istanbul University Faculty of Medicine (IU)&74&73\\
Erasmus Medical Center (EMC)&91&103\\
\hline
\textbf{Total}&\textbf{723}&\textbf{738}\\
\hline
\end{tabular}
\label{tab: Distribution}
\end{center}
\end{table}

\begin{table*}[tb]
\setlength{\tabcolsep}{3pt}
\caption{Pancreas MRI segmentation results. ``AAW'' denotes our proposed Adaptive Aggregation Weights method. Improved results are highlighted in pink, while degraded results are highlighted in cyan.}
\begin{center}
\begin{tabular}{l|cccccc|cccccc}
\hline
&\multicolumn{6}{c|}{\textbf{T1-Weighted Modality}}&\multicolumn{6}{c}{\textbf{T2-Weighted Modality}}\\
\hline
\textbf{Method}&\textbf{Dice$\uparrow$}  & \textbf{Jaccard$\uparrow$}  & \textbf{Precision$\uparrow$}& \textbf{Recall$\uparrow$} & \textbf{HD95$\downarrow$} & \textbf{ASSD$\downarrow$}&\textbf{Dice$\uparrow$}  & \textbf{Jaccard$\uparrow$}  & \textbf{Precision$\uparrow$}& \textbf{Recall$\uparrow$} & \textbf{HD95$\downarrow$} & \textbf{ASSD$\downarrow$}\\
\hline
\multicolumn{13}{l}{\textbf{Center 1: New York University Langone Health (NYU)}}\\
\hline
No FL&0.8194 & 0.7065 & 0.8502 & 0.7979 & 2.3686 & 0.4106&0.7863 & 0.6763 & 0.8200 & 0.7690 & 3.9018 & 0.7015\\
FedAvg~\cite{mcmahan2017communication}& 0.6839 & 0.5310 & 0.6071 & 0.8096 & 7.1474 & 1.4332&0.6515 & 0.5064 & 0.6071 & 0.7498 & 9.5187 & 2.1262\\
\textbf{FedAvg+AAW}&\cellcolor{Better} 0.7133 & \cellcolor{Better}0.5709 & \cellcolor{Better}0.6269 & \cellcolor{Better}0.8447 & \cellcolor{Worse}8.1295 & \cellcolor{Worse}1.9601&\cellcolor{Worse}0.5812 & \cellcolor{Worse}0.4342 & \cellcolor{Worse}0.5534 & \cellcolor{Worse}0.6657 & \cellcolor{Worse}12.9598 & \cellcolor{Worse}3.5244\\
\hline
\multicolumn{13}{l}{\textbf{Center 2: Mayo Clinic Florida (MCF)}}\\
\hline
No FL&0.8279 & 0.7226 & 0.8367 & 0.8287 & 3.4076 & 0.6147&0.7993 & 0.7089 & 0.8735 & 0.7762 & 3.9265 & 1.0606\\
FedAvg~\cite{mcmahan2017communication}&0.6657 & 0.5183 & 0.6649 & 0.7393 & 10.6307 & 2.2033&0.5888 & 0.4811 & 0.7825 & 0.5628 & 17.9339 & 4.9035\\
\textbf{FedAvg+AAW}&\cellcolor{Better}0.7510 & \cellcolor{Better}0.6222 & \cellcolor{Better}0.6917 & \cellcolor{Better}0.8613 & \cellcolor{Better}7.2299 & \cellcolor{Better}1.6896&\cellcolor{Better}0.6090 & \cellcolor{Better}0.4829 & \cellcolor{Worse}0.7356 & \cellcolor{Better}0.6031 & \cellcolor{Better}15.9095 & \cellcolor{Better}3.6724\\
\hline
\multicolumn{13}{l}{\textbf{Center 3: Northwestern University (NU)}}\\
\hline
No FL&0.8184 & 0.7205 & 0.8443 & 0.8056 & 4.4057 & 1.5844&0.8417 & 0.7595 & 0.8717 & 0.8306 & 2.2739 & 0.5431\\
FedAvg~\cite{mcmahan2017communication}&0.6578 & 0.5045 & 0.5181 & 0.9145 & 4.6909 & 1.2883&0.6785 & 0.5416 & 0.5917 & 0.8185 & 4.4509 & 1.2742\\
\textbf{FedAvg+AAW}&\cellcolor{Better}0.6736 & \cellcolor{Better}0.5241 & \cellcolor{Better}0.5536 & \cellcolor{Worse}0.8752 &\cellcolor{Worse}5.2923 & \cellcolor{Worse}1.4514&\cellcolor{Worse}0.6150 & \cellcolor{Worse}0.4688 & \cellcolor{Worse}0.5354 & \cellcolor{Worse}0.7512 & \cellcolor{Worse}5.2034 & \cellcolor{Worse}1.5686\\
\hline
\multicolumn{13}{l}{\textbf{Center 4: Allegheny Health Network (AHN)}}\\
\hline
No FL&0.8321 & 0.7179 & 0.8549 & 0.8129 & 1.7661 & 0.3060&0.6492 & 0.5306 & 0.7420 & 0.5996 & 8.1092 & 1.8968\\
FedAvg~\cite{mcmahan2017communication}&0.4117 & 0.2727 & 0.6065 & 0.4587 & 24.3632 & 7.2514&0.3664 & 0.2376 & 0.5362 & 0.3190 & 25.4220 & 6.8224\\
\textbf{FedAvg+AAW}&\cellcolor{Better}0.6953 & \cellcolor{Better}0.5568 & \cellcolor{Better}0.7053 & \cellcolor{Better}0.7997 & \cellcolor{Better}5.1412 & \cellcolor{Better}1.0850&\cellcolor{Better}0.4916 &\cellcolor{Better}0.3380 &\cellcolor{Better}0.5650 & \cellcolor{Better}0.4672 & \cellcolor{Better}12.1286 & \cellcolor{Better}2.8360\\
\hline
\multicolumn{13}{l}{\textbf{Center 5: Mayo Clinic Arizona (MCA)}}\\
\hline
No FL&0.7674 & 0.6471 & 0.7686 & 0.7682 & 2.3627 & 0.5067&0.6059 & 0.5003 & 0.6336 & 0.5969 & 10.2899 & 2.3163\\
FedAvg~\cite{mcmahan2017communication}&0.5975 & 0.4453 & 0.4937 & 0.7779 & 8.4508 & 1.9823&0.4077 & 0.3112 & 0.4278 & 0.3943 & 13.8540 & 4.0907\\
\textbf{FedAvg+AAW}&\cellcolor{Better}0.6680 &\cellcolor{Better}0.5121 & \cellcolor{Better}0.5597 &\cellcolor{Better}0.8386 & \cellcolor{Better}3.9422 & \cellcolor{Better}0.9641&\cellcolor{Better}0.4282 &\cellcolor{Better}0.3167 & \cellcolor{Worse}0.3717 &\cellcolor{Better}0.5103 &\cellcolor{Better}8.9490 & \cellcolor{Better}3.1265\\
\hline
\multicolumn{13}{l}{\textbf{Center 6: Istanbul University Faculty of Medicine (IU)}}\\
\hline
No FL&0.8394 & 0.7454 & 0.8455 & 0.8344 & 2.7811 & 0.4969&0.8360 & 0.7578 & 0.8292 & 0.8654 & 3.7258 & 0.8394\\
FedAvg~\cite{mcmahan2017communication}&0.6277 & 0.4716 & 0.4844 & 0.9122 & 8.2974 & 1.7792&0.7179 & 0.5859 & 0.6186 & 0.8681 & 7.1870 & 1.3420\\
\textbf{FedAvg+AAW}&\cellcolor{Better}0.6564 & \cellcolor{Better}0.5038 & \cellcolor{Better}0.5418 & \cellcolor{Worse}0.8499 &\cellcolor{Better} 6.2866 & \cellcolor{Better}1.3795&\cellcolor{Worse}0.5506 & \cellcolor{Worse}0.3994 &\cellcolor{Worse}0.5100 & \cellcolor{Worse}0.6396 & \cellcolor{Worse}9.7447 &\cellcolor{Worse}2.5334\\
\hline
\multicolumn{13}{l}{\textbf{Center 7: Erasmus Medical Center (EMC)}}\\
\hline
No FL&0.8263 & 0.7402 & 0.9124 & 0.8021 & 4.0791 & 1.5077&0.8107 & 0.7350 & 0.8626 & 0.7987 & 3.9553 & 1.1469\\
FedAvg~\cite{mcmahan2017communication}&0.6952 & 0.5398 & 0.5762 & 0.9010 & 8.2871 & 1.2091& 0.6387 & 0.5155 & 0.5821 & 0.7362 & 11.0488 & 3.1895\\
\textbf{FedAvg+AAW}& \cellcolor{Better}0.7121 & \cellcolor{Better}0.5743 &\cellcolor{Better} 0.6593 & \cellcolor{Worse}0.8411 & \cellcolor{Better}4.9892 & \cellcolor{Worse}1.2401&\cellcolor{Worse}0.6012 & \cellcolor{Worse}0.4679 & \cellcolor{Worse}0.5259 &\cellcolor{Better}0.7441 & \cellcolor{Better}8.1523 & \cellcolor{Better}2.2427\\
\hline
\multicolumn{13}{l}{\textbf{Average}}\\
\hline
No FL&0.8223 & 0.7203 & 0.8507 & 0.8100 & 3.4022 & 0.9351&0.8011 & 0.7103 & 0.8422 & 0.7888 & 3.8224 & 0.9030\\
FedAvg~\cite{mcmahan2017communication}&0.6583 & 0.5061 & 0.5733 & 0.8367 & 7.9286 & 1.7303& 0.6324 & 0.5034 & 0.6254 & 0.7140 & 10.4819 & 2.7442\\
\textbf{FedAvg+AAW}&\cellcolor{Better}0.7017 & \cellcolor{Better}0.5593 & \cellcolor{Better}0.6146 & \cellcolor{Better}0.8554 & \cellcolor{Better}6.3298 & \cellcolor{Better}1.5523&\cellcolor{Worse}0.5870 & \cellcolor{Worse}0.4467 & \cellcolor{Worse}0.5697 &\cellcolor{Worse}0.6727 &\cellcolor{Better}10.2151 & \cellcolor{Better}2.6916\\
\hline
\end{tabular}\vspace{-10pt}
\label{tab: fed}
\end{center}
\end{table*}

SynergyNet~\cite{gorade2024synergynet} has achieved state-of-the-art performance in various 2D medical image segmentation tasks. In this study, we extended SynergyNet-8s2h to a 3D version by replacing its 2D convolutional blocks with 3D convolutional blocks, making it suitable for pancreas MRI segmentation. Other approaches can be adapted to our framework once they are shown suitable for pancreas segmentation. The training process employed the AdamW optimizer~\cite{loshchilov2017decoupled} with a learning rate of 1e-4, a batch size of 8, and a weight decay of 0.01. We trained the model for 300 epochs to minimize the Dice-BCE loss $\mathcal{L}$:
\begin{equation}
    \mathcal{L}_{Dice} = 1 - \frac{2 \cdot |\mathbf{Y} \cap \hat{\mathbf{Y}}|}{|\mathbf{Y}| + |\hat{\mathbf{Y}}|} = 1 - \frac{2 \cdot \sum_{i=0}^{N-1} y_i \cdot \hat{y}_i}{\sum_{i=0}^{N-1} y_i + \sum_{i=0}^{N-1} \hat{y}_i},
\end{equation}
\begin{equation}
    \mathcal{L}_{BCE} = -\frac{1}{N} \sum_{i=0}^{N-1} \left[ y_i  \log(\hat{y}_i) + (1 - y_i)  \log(1 - \hat{y}_i) \right],
\end{equation}
\begin{equation}
    \mathcal{L} = \mathcal{L}_{Dice} + \mathcal{L}_{BCE},
\end{equation}
where, $\mathbf{Y}$ and $\hat{\mathbf{Y}}$ stand for the true and the predicted segmentation masks with $N$ pixels, respectively. $y_i$ and $\hat{y}_i$ denote the $i$-th pixels of $\mathbf{Y}$ and $\hat{\mathbf{Y}}$. All images were resized to $80\times256\times256$ for training and testing.
To compare the performance of different methods, we evaluated the models based on the dice coefficient, Jaccard index (intersection over union), precision, recall, Hausdorff distance (HD95), and average symmetric surface distance (ASSD).
The lower HD95 and ASSD and the higher other metrics indicate better performance.

Table~\ref{tab: fed} presents a comprehensive analysis of pancreas MRI segmentation performance across multiple client centers using the proposed adaptive aggregation weights strategy within an FL framework. The reported average results are weighted based on the number of test samples from each center, ensuring that larger datasets have a proportionate influence on the overall performance metrics. For benchmarking purposes, we also include segmentation results obtained without the use of FL, which serves as an upper-bound reference. We observed that on the T1 modality, the adaptive aggregation weights significantly improve the FedAvg results with region-based metrics such as Dice and Jaccard. On the T2 modality, we observed reduced HD95 and ASSD metrics while did not observe much improvement from region-based metrics. We also observed, naturally, that some centers underperformed in the FL setting. This was expected too because of the number of images they contributed was small and quality varies significantly. Most of the results were better than conventional FedAvg, and approaching no-FL results. 
\vspace{-10pt}
\section{Conclusion}
This study presents an innovative adaptive aggregation weights strategy tailored for FL frameworks, specifically addressing the complexities of pancreas MRI segmentation. Traditional FL approaches aggregate model updates uniformly across clients, which can be suboptimal in heterogeneous clinical environments where local data distributions vary significantly. Our proposed method dynamically adjusts the contribution of each client during the model aggregation phase based on intrinsic local data characteristics, thereby mitigating the adverse effects of domain shift and enhancing model generalization across diverse clinical centers. This targeted aggregation approach not only improves the robustness of the segmentation model but also maintains the integrity and privacy of patient data, as raw data remains localized and secure within each institution.

Experimental validation was conducted using multi-center MRI datasets (of the pancreas) encompassing various imaging protocols and patient populations. Our results demonstrate a significant enhancement in segmentation performance metrics, notably a reduction in the HD95 and the ASSD. These improvements indicate a finer anatomical boundary delineation and more consistent segmentation accuracy across varying datasets. Moreover, the adaptive aggregation strategy effectively addresses the variability inherent in FL environments, ensuring that the global model remains resilient and generalizable without compromising patient privacy.


\section{Compliance with ethical standards}
\label{sec:ethics}
This study was performed in line with the principles of the Declaration of Helsinki. Approval was granted by Northwestern University (No. STU00214545).

\section{Acknowledgments}
\label{sec:acknowledgments}
This work was supported by the following grants: NIH NCI R01-CA246704, R01-CA240639, U01-CA268808, NIH/NHLBI R01-HL171376, and NIH/NIDDK \#U01 DK127384-02S1. 

\bibliographystyle{IEEEbib}
\bibliography{strings,refs}

\end{document}